\documentclass[letterpaper,aps,prb,twocolumn,nofootinbib,superscriptaddress]{revtex4}

\usepackage{amsmath}
\usepackage{bm}
\usepackage{graphicx}
\usepackage[colorlinks=true,citecolor=blue,linkcolor=blue]{hyperref}

\usepackage[T1]{fontenc}
\usepackage{txfonts}

\begin{document}

\title{Vibrational Excitations in Weakly Coupled Single-Molecule Junctions: A Computational Analysis}

\author{Johannes S. Seldenthuis}
\author{Herre S. J. van der Zant\footnote{Address correspondence to h.s.j.vanderzant@tudelft.nl.}}
\affiliation{Kavli Institute of Nanoscience, Delft University of Technology, Lorentzweg 1, 2628 CJ Delft}
\author{Mark A. Ratner}
\affiliation{Department of Chemistry, Northwestern University, Evanston, Illinois 60208}
\author{Joseph M. Thijssen}
\affiliation{Kavli Institute of Nanoscience, Delft University of Technology, Lorentzweg 1, 2628 CJ Delft}

\date{\today}

\begin{abstract}
In bulk systems, molecules are routinely identified by their vibrational spectrum using Raman or infrared spectroscopy. In recent years, vibrational excitation lines have been observed in low-temperature conductance measurements on single molecule junctions and they can provide a similar means of identification. We present a method to efficiently calculate these excitation lines in weakly coupled, gateable single-molecule junctions, using a combination of \emph{ab initio} density functional theory and rate equations. Our method takes transitions from excited to excited vibrational state into account by evaluating the Franck-Condon factors for an arbitrary number of vibrational quanta, and is therefore able to predict qualitatively different behaviour from calculations limited to transitions from ground state to excited vibrational state. We find that the vibrational spectrum is sensitive to the molecular contact geometry and the charge state, and that it is generally necessary to take more than one vibrational quantum into account. Quantitative comparison to previously reported measurements on $pi$-conjugated molecules reveals that our method is able to characterize the vibrational excitations and can be used to identify single molecules in a junction. The method is computationally feasible on commodity hardware.
\end{abstract}

\maketitle

In recent years, vibrational excitations of single molecules have been measured in the inelastic tunneling regime with scanning tunneling microscopes (STMs),~\cite{stipe} in mechanical breakjunctions (MBJs)~\cite{smit,parks} and in the sequential tunneling regime (SET) with electromigrated breakjunctions (EBJs).~\cite{park,yu,pasupathy} Measurements in EBJs on an oligophenylenevinylene derivative by Osorio~\emph{et al.}~\cite{osorio} reveal a vibrational spectrum of 17 excitations in the sequential tunneling regime (see figure~\ref{fig:opv5}a). It has been shown that these modes are consistent with Raman (above 15~meV) and infrared (above 50~meV) spectroscopy data. However, the Raman and IR data show more peaks than are observed in the transport measurement. Moreover, the Raman and IR measurements were performed on polycrystalline samples and KBr pellets respectively, which do not reflect the conditions of the molecule in the junction.

Theoretical investigations on vibrational excitations in the SET regime have so far mainly concentrated on small systems with only one vibrational mode~\cite{boese,braig,mitra,wegewijs,koch2,koch1}. Chang~\emph{et al.}~\cite{chang} use density functional theory (DFT) calculations on the C$_{72}$ fullerene dimer to calculate all vibrational modes, restricting themselves to transitions from the ground state in one charge state to the vibrational excited state of the other charge state. We have developed a computationally efficient method to calculate the vibrational spectrum of a sizeable molecule in the sequential tunneling regime, based on first principles DFT calculations to obtain the vibrational modes in a three-terminal setup. This method takes the charge state and contact geometry of the molecule into account and predicts the relative intensities of vibrational excitations. In addition, transitions from excited to excited vibrational state are accounted for by evaluating the Franck-Condon factors involving several vibrational quanta. Our method can therefore predict qualitatively different behaviour compared to calculations that only include transitions from ground state to excited vibrational state.~\cite{koch2}

\begin{figure*}
    \begin{center}
        \includegraphics[width=0.975\textwidth]{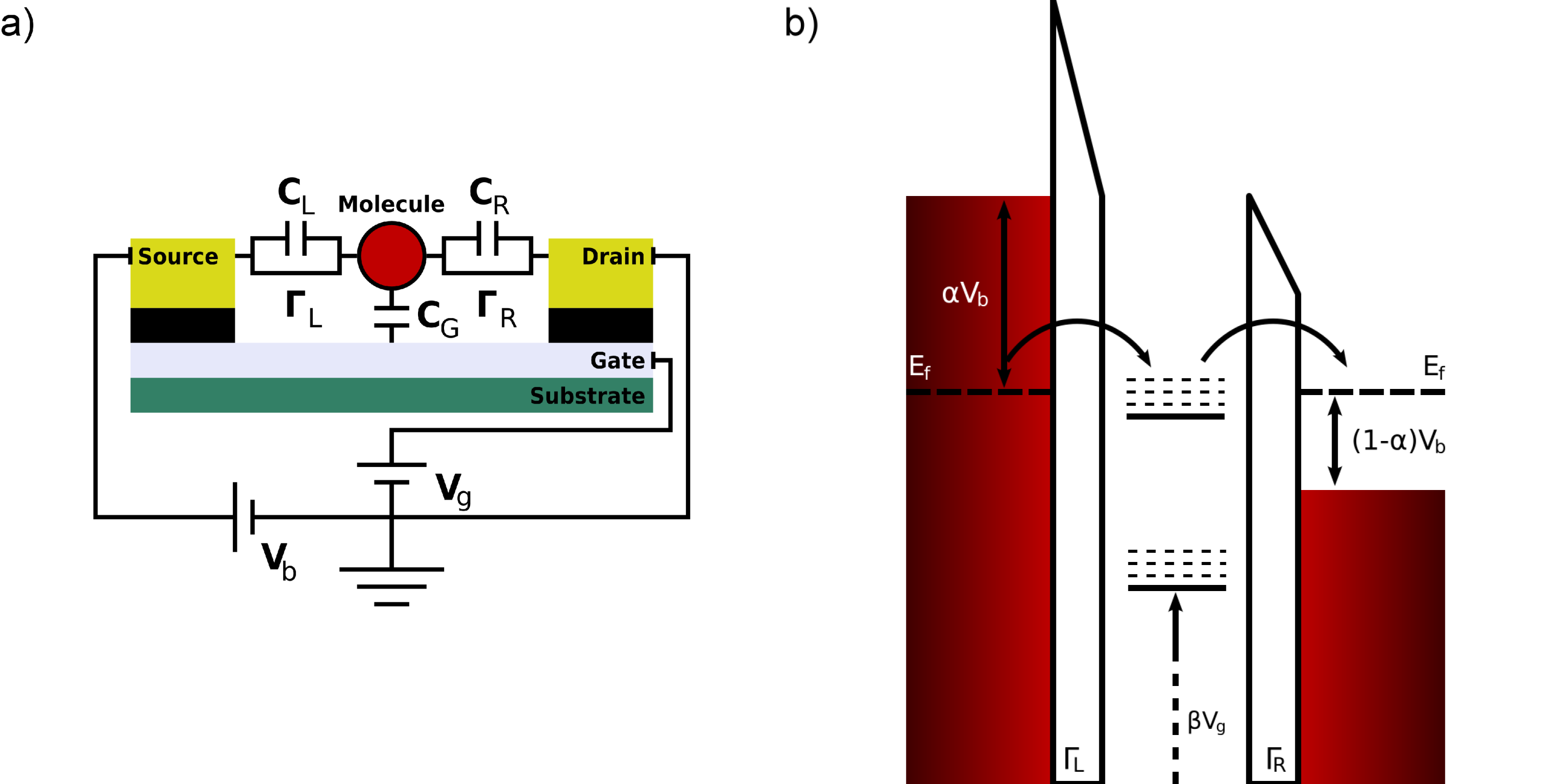}
    \end{center}
    \caption{\label{fig:device} (a) Schematic picture of a gateable junction containing a single molecule capacitively coupled to the left ($C_L$), right ($C_R$) and gate ($C_G$) electrode. (b) The transport mechanism in the sequential tunneling regime. The distribution of the bias voltage over the leads is given by $\alpha=\frac{C_R+\frac{1}{2}C_G}{C_L+C_R+C_G}$ and the gate coupling by $\beta=\frac{C_G}{C_L+C_R+C_G}$.~\cite{bonet}}
\end{figure*}

A schematic picture of a gateable electromigrated breakjunction containing a single molecule is shown in figure~\ref{fig:device}a. The couplings to the leads are given by $\Gamma_L$ and $\Gamma_R$. In the weak coupling limit, $\Gamma_L,\Gamma_R,kT\ll\Delta E,E_C$, the level spacing ($\Delta E$) and charging energy ($E_C$) of the molecule allow only one electron to tunnel onto the molecule at a time (sequential tunneling). The transport mechanism for this case is shown in figure~\ref{fig:device}b. While tunneling on or off a molecule, an electron can excite a vibrational mode, which may show up as a line running parallel to the diamond edges in the stability diagram (a plot of the conductance as a function of bias and gate voltage).~\cite{thijssen} In this paper we calculate such stability diagrams with a rate equation approach and present a comparison with experimental results.

\section*{Results and Discussion}

\begin{figure*}
    \begin{center}
        \includegraphics[width=0.975\textwidth]{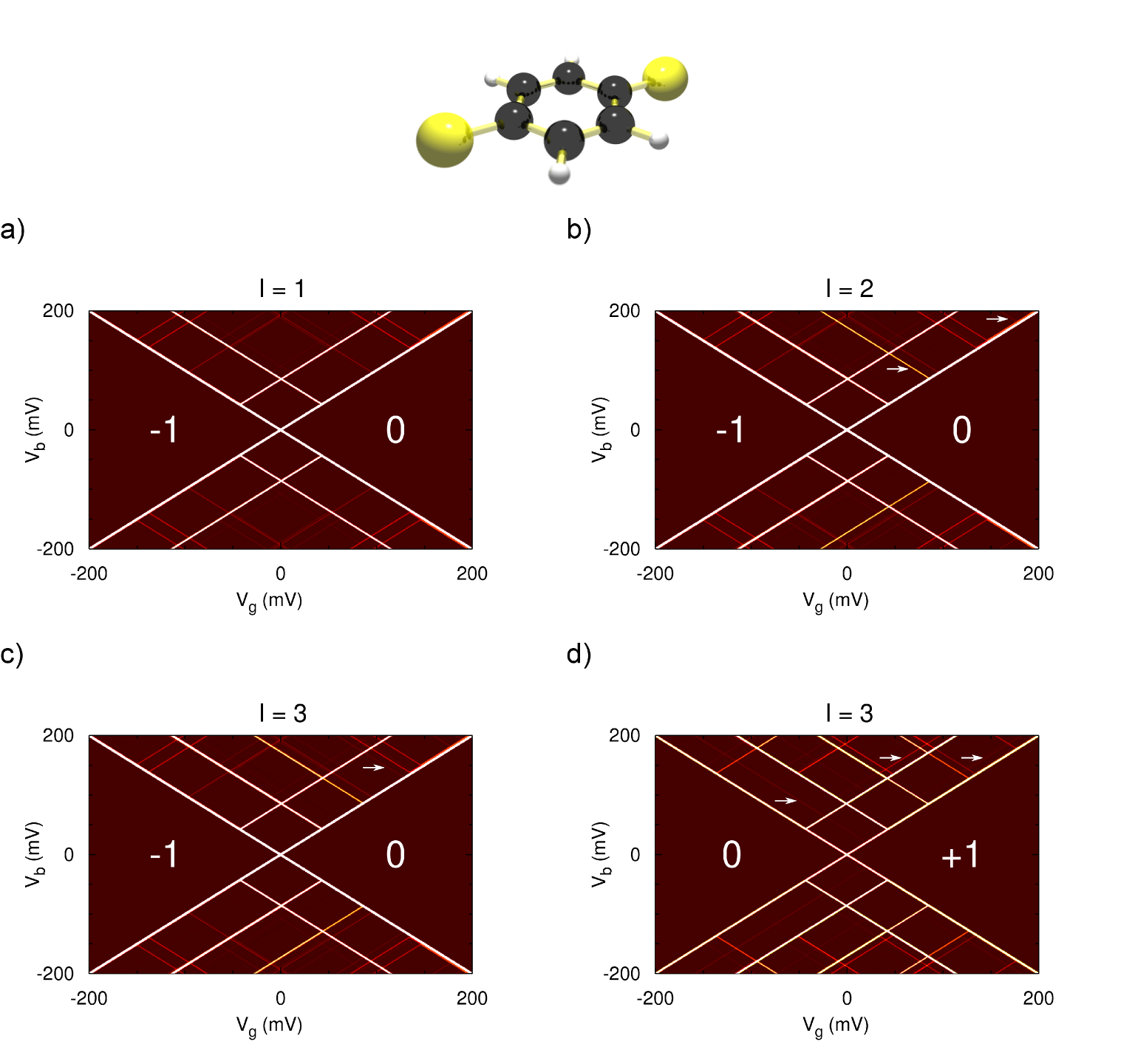}
    \end{center}
    \caption{\label{fig:bdt} Calculated stability diagrams of the -1$\rightarrow$0 (a--c) and 0$\rightarrow$+1 (d) transitions in benzenedithiol with increasing number of vibrational quanta. The arrows point to the main differences between the diagrams (see text). Since the calculation is symmetric in the bias voltage, they are only shown for positive bias.}
\end{figure*}

\begin{figure*}
    \begin{center}
        \includegraphics[width=0.975\textwidth]{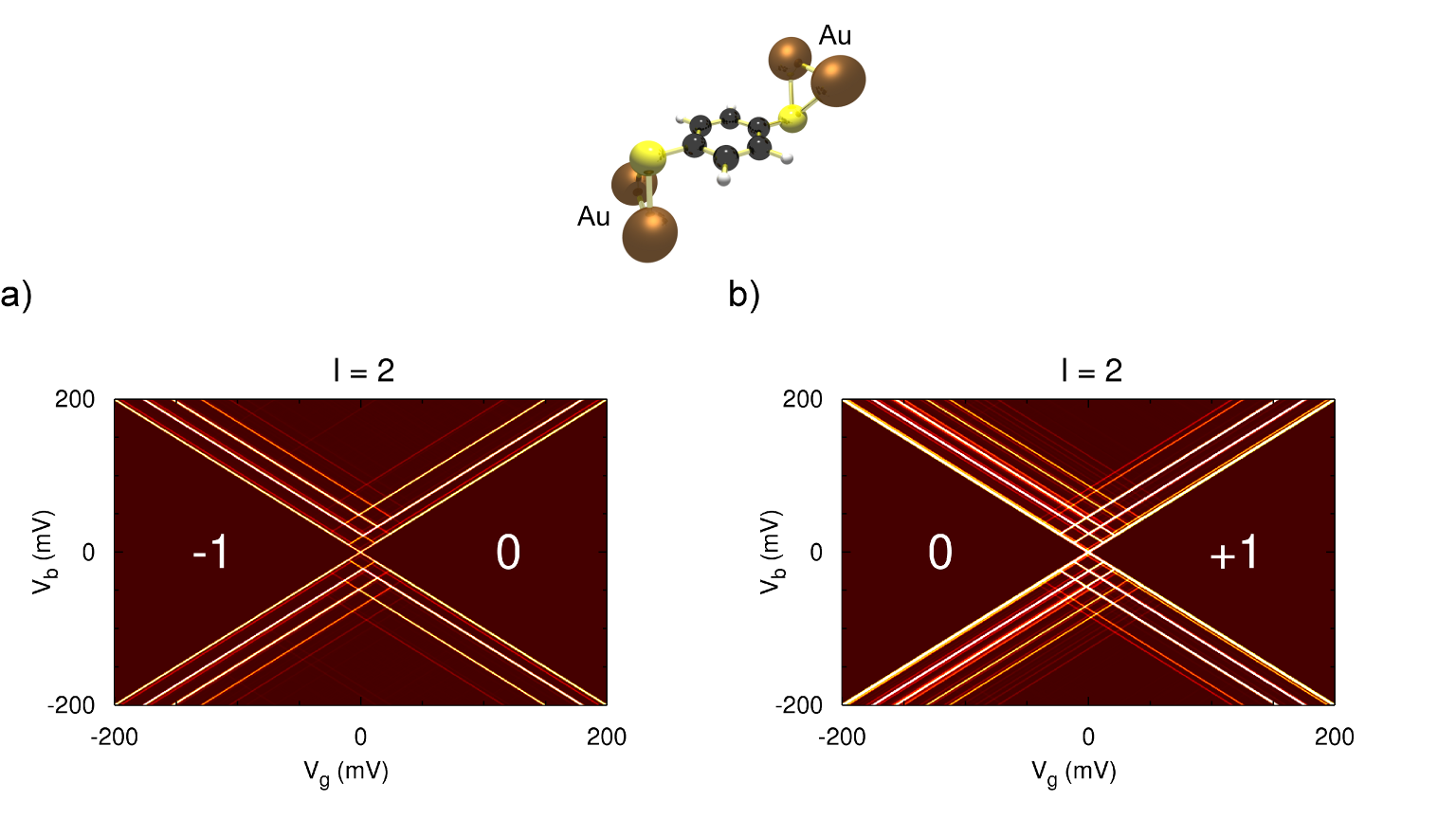}
    \end{center}
    \caption{\label{fig:bdtau} Calculated stability diagrams of the -1$\rightarrow$0 (a) and 0$\rightarrow$+1 (b) transition in benzenedithiol with two gold atoms on either side to simulate the leads. Two vibrational quanta are taken into account.}
\end{figure*}

\begin{figure*}
    \begin{center}
        \includegraphics[width=0.975\textwidth]{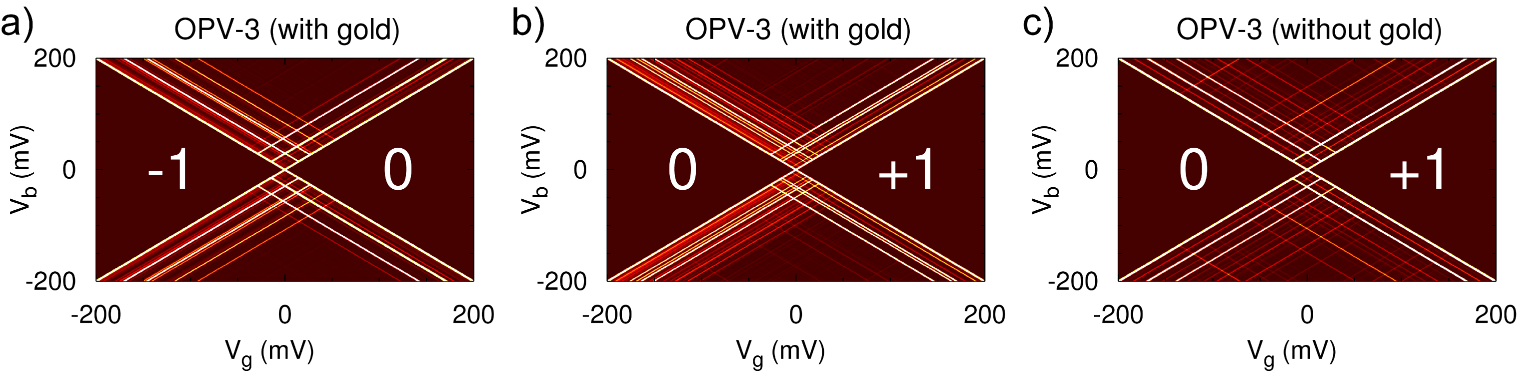}
    \end{center}
    \caption{\label{fig:opv3} Calculated stability diagrams for OPV-3 with (a and b) and without (c) gold for two charge state transitions. The calculations take two vibrational quanta into account.}
\end{figure*}

\begin{figure*}
    \begin{center}
        \includegraphics[width=0.975\textwidth]{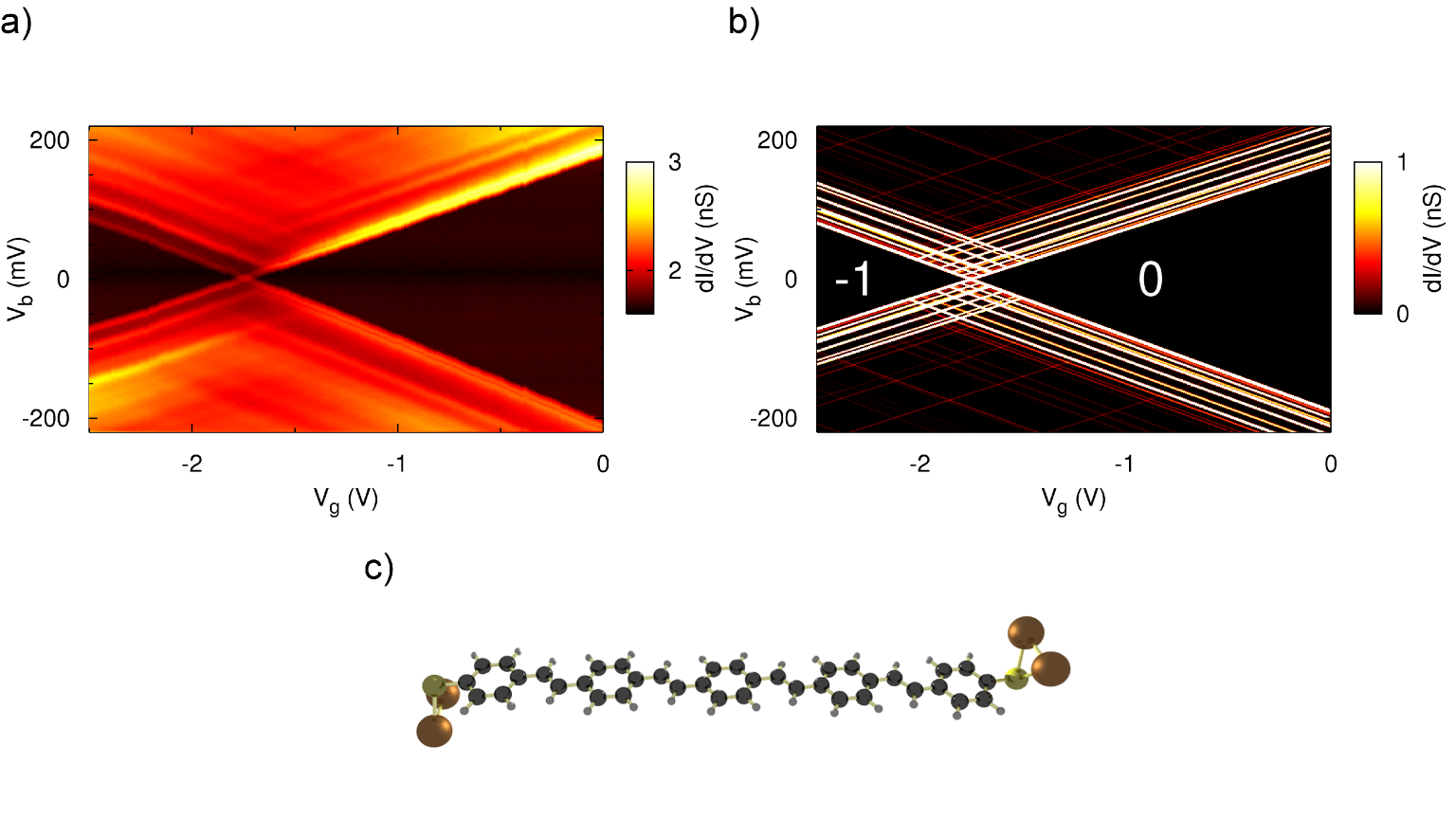}
    \end{center}
    \caption{\label{fig:opv5} (a) Measured stability diagram of a junction containing OPV-5.~\cite{osorio} (b) Calculated stability diagram of an OPV-5 molecule with one vibrational quantum. (c) The configuration of the OPV-5 molecule in the calculations. The dodecane sidearms of the measured molecule (see figure~1a in \cite{osorio}) are omitted and two gold atoms are added on either side to simulate the leads.}
\end{figure*}

We have applied our rate equation method to three molecules with increasing length: benzenedithiol (see figures \ref{fig:bdt} and \ref{fig:bdtau}), the oligophenylenevinylene derivative OPV-3 (see figure~\ref{fig:opv3}) and OPV-5 (see figure~\ref{fig:opv5}). All vibrational mode calculations are carried out using the Amsterdam Density Functional code~\cite{adf1,adf2} with the Analytical Second Derivatives module.~\cite{adf3}

As a first example, we present our results for benzenedithiol adsorbed on gold. This example is used in order to demonstrate our method; In experiments, this system is generally not weakly coupled. We test the method by studying the influence of the number of vibrational quanta, the charge state and the presence of gold contacts on the stability diagrams. The stability diagrams are calculated with a symmetric coupling to the leads of 1~meV, a bias ($\alpha$) and gate ($\beta$) coupling of 0.5 and a temperature of 1.6~K. The resulting stability diagram for the -1$\rightarrow$0 transition in bare benzenedithiol with one vibrational quantum (961 Franck-Condon factors) is shown in figure~\ref{fig:bdt}a. Of the 25 vibrational modes with energies below 200~meV, eight excitation lines are visible belonging to the -1 state, and seven belonging to the neutral state. For both states there are three excitation lines that do not continue all the way to the diamond edge, but stop at the strong line at 43~meV.

Taking two vibrational quanta into account (see figure~\ref{fig:bdt}b), reveals two new excitation lines for the neutral state, a strong line at 86~meV at a weak line at 180~meV (see the white arrows). Taking one more quantum into account (see figure~\ref{fig:bdt}c) adds a weak excitation line for the neutral charge state at 129~meV, again indicated by a white arrow. For the 0$\rightarrow$+1 transition (figure~\ref{fig:bdt}d), small changes are found. The arrows point to excitation lines that are absent in figure~\ref{fig:bdt}c. Compared to that diagram, the higher energy excitations (above 100~meV) have shifted by several meV.

In a junction, the molecule is bonded to the gold contacts. We have modeled this by adding two gold atoms on either side of the molecule. The resulting stability diagrams (with $l=2$) are shown in figure~\ref{fig:bdtau}. These diagrams are quite different from those of the same charge state transitions in figure~\ref{fig:bdt}c and d. Depending on the charge state, five to eight excitations are visible below 75~meV, but no higher modes are observed. The electron-phonon couplings for the neutral charge state in the transition of figure~\ref{fig:bdtau}a are shown in figure~\ref{fig:eph}a. Two modes have a large electron-phonon coupling (with coupling strengths larger than 1), showing that it is necessary for this system to take more than one vibrational quantum into account.~\cite{koch2}

The calculations show that only a few of the 30 vibrational modes of benzenedithiol are expected to be visible in transport measurements and that they are dependent on the charge state and sensitive to the contact geometry. For some modes in this molecule it is necessary to take more than one vibrational quantum into account. For example, the modes at 86~meV (for $l=2$) and 129~meV (for $l=3$) are probably higher harmonics of the strong excitation at 43~meV. The fact that several other lines stop at this excitation shows that it is also necessary to take the Franck-Condon factors for excited vibrational state to excited vibrational state into account.

The second molecule for which we have calculated the vibrational spectrum is OPV-3. As with benzenedithiol, the gold contacts are simulated by adding two gold atoms on either side of the molecule. The results for two charge state transitions with gold and one without are shown in figure~\ref{fig:opv3}. The calculations take two vibrational quanta into account. Comparing the calculations to those on benzenedithiol indicates that OPV-3 is less sensitive to the contact geometry. OPV-3 without gold has more modes at lower energies than benzenedithiol and the modes at higher energies are much less suppressed when the two gold atoms are added. Also, the electron-phonon couplings for OPV-3 are smaller than for benzenedithiol (see figure~\ref{fig:eph}b). These trends are not unexpected since OPV-3 is a larger molecule and the atoms will on average be further away from the leads, leading to a smaller sensitivity to the contact geometry. Also, since OPV-3 is conjugated, an extra electron will be delocalized over the entire molecule, and the atomic displacements will be smaller, resulting in a smaller electron-phonon coupling. In the case of OPV-3 we have performed several calculations with different contact geometries. We find that adding up to 19 gold atoms on either side of the molecule has no significant effect on the vibrational modes above 20~meV.~\footnote{We have also performed measurements on vibrational excitations in OPV-3. However, broadening of the lines due to large couplings to the leads prevents us from obtaining measurements with sufficient resolution to make a quantitative comparison to the calculations possible. The measurements do show the same trends as the calculations. None of the samples show any excitations above 30~meV, and only a few below.}

\begin{figure*}
    \begin{center}
        \includegraphics[width=0.975\textwidth]{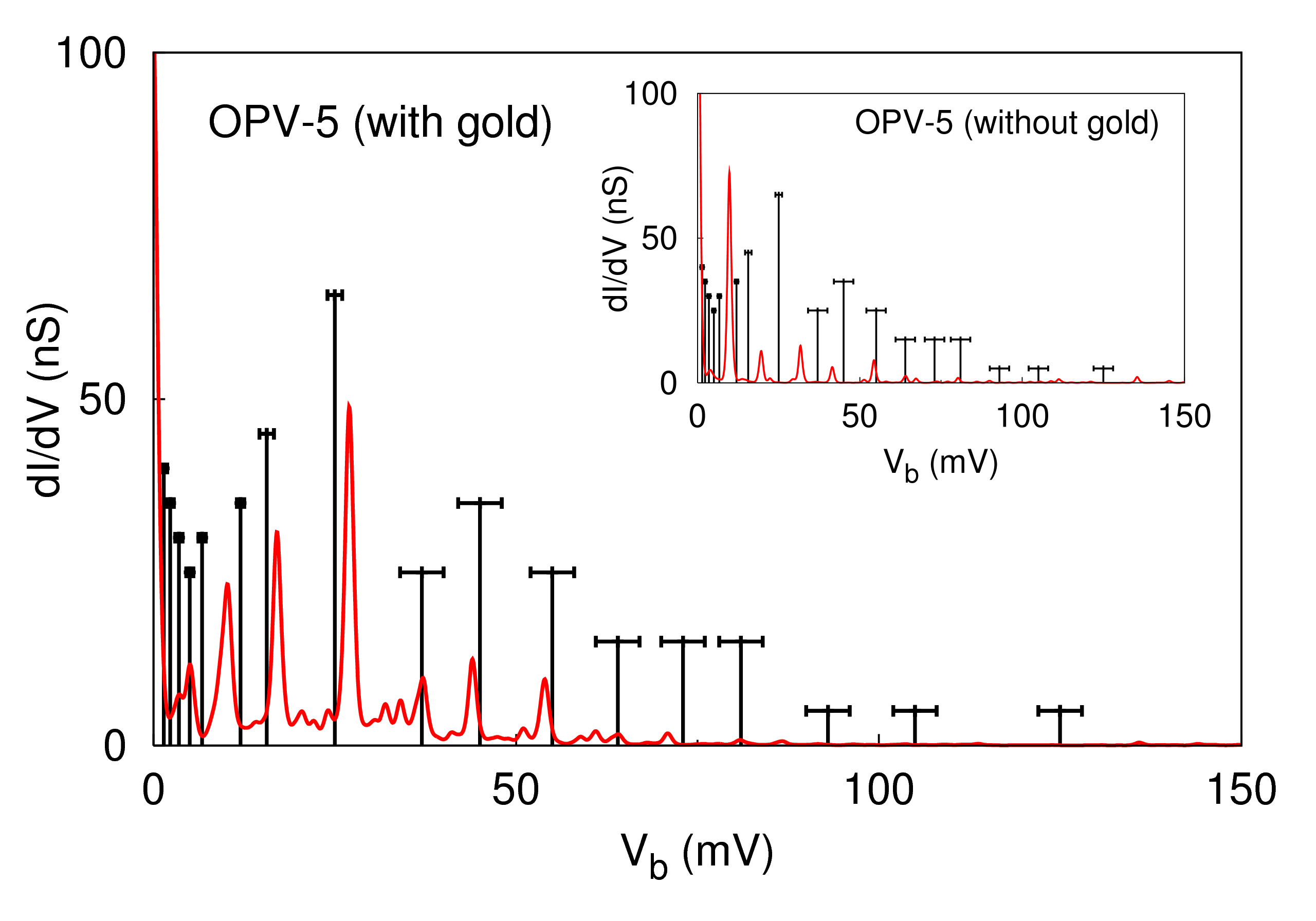}
    \end{center}
    \caption{\label{fig:trace} Calculated $dI/dV$-trace of the diamond edge of the neutral charge state in figure~\ref{fig:opv5}b, taking three vibrational quanta into account. The inset shows the same calculations, but with the gold atoms omitted. All measured excitations in this energy range (see figure~\ref{fig:opv5}a) are shown. The uncertainties in the measured energies are indicated by the horizontal bars.}
\end{figure*}

\begin{figure*}
    \begin{center}
        \includegraphics[width=0.975\textwidth]{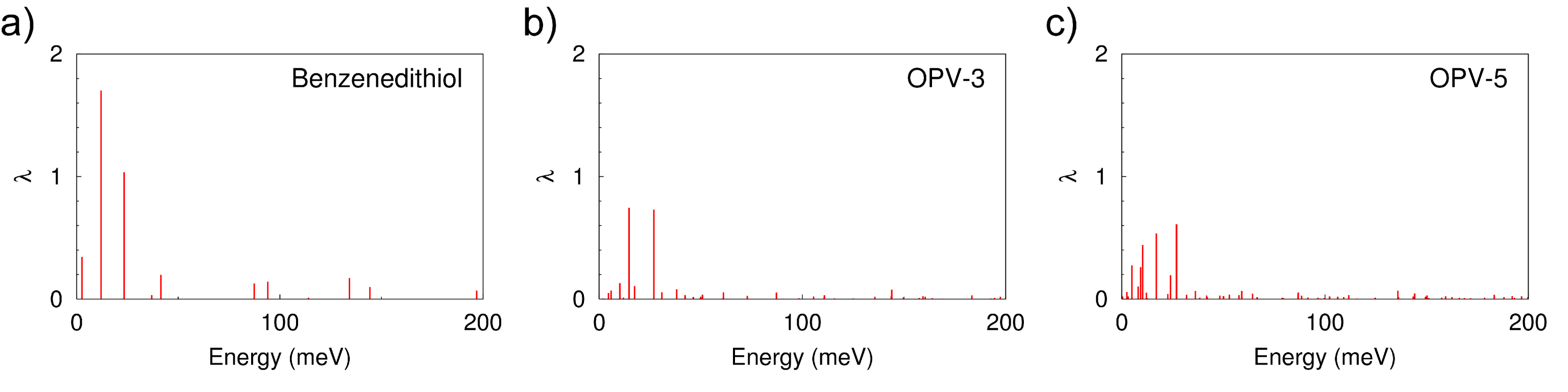}
    \end{center}
    \caption{\label{fig:eph} Dimensionless electron-phonon couplings ($\lambda$) for the vibrational modes of the neutral charge state at the $-1\rightarrow 0$ transition for (a) benzenedithiol, (b) OPV-3 and (c) OPV-5. All calculations include two gold atoms on either side of the molecule to simulate the leads. Analysis of the atomic displacements shows that primarily modes that distort the $\pi-\pi$ overlap give rise to a non-zero electron-phonon coupling.}
\end{figure*}

The calculated stability diagram of the third molecule, OPV-5, is shown in figure~\ref{fig:opv5}b. The temperature and coupling parameters of the calculation are chosen to be the same as in the experiment (figure~\ref{fig:opv5}a). Although we are unable to determine the charge states in the measurement, the fact that the degeneracy point is the first at a negative gate voltage suggests a -1$\rightarrow$0 transition (see also figure~3 in \cite{osorio}). In the calculation, the non-conjugated dodecane sidearms of the measured molecule are omitted. These arms are  not expected to influence the electronic transport and will most likely only affect the low-energy vibrational modes. As with benzenedithiol, the contacts are modelled by adding two gold atoms on either side of the molecule. The calculation takes one vibrational quantum into account.

Figure~\ref{fig:trace} shows a $dI/dV$-trace along the diamond edge of the neutral charge state in figure~\ref{fig:opv5}b. However, since this is a smaller calculation, three vibrational quanta can be taken into account. The peaks in this figure correspond to the excitation lines in the calculated stability diagram. In the experimental stability diagram, a background conductance makes it difficult to resolve all excitation lines at the same color scale, but close inspection reveals 17 modes in the energy range below 125~meV (see figure~\ref{fig:trace} and table~I in \cite{osorio}). The energies of the excitations in the measurement are determined from the bias voltage at which they cross the diamond edge. Broadening due to the temperature and the leads introduces an uncertainty, indicated by the horizontal bars in figure~\ref{fig:trace}.

Figure~\ref{fig:trace} reveals a close match between the experiment and the calculation for the modes between 10 and 80~meV. The calculation shows several small peaks in this range not observed in the measurement. It should be noted that in the rate equation approach, broadening of excitation lines is solely due to temperature. Broadening due to the couplings to the leads is not accounted for. Calculations which do take this broadening into account show that these small peaks are smeared out and the calculation and measurement show the same number of peaks in the aforementioned range.

The inset in figure~\ref{fig:trace} shows the same calculation, but with the gold atoms omitted. Comparison with the measurement shows a large discrepancy for excitations below 50~meV. It is clear from this figure that the addition of two gold atoms on either side of the molecule can already account for most of the influence of the contact geometry on modes above 10~meV. The charging energy of an OPV-5 molecule in a junction is an order of magnitude smaller than the difference between the ionization energy and electron affinity of the molecule in the gas phase,\cite{kubatkin} probably due to screening in the leads. While this effect changes the energies of the orbitals, the shape of the orbitals, and therefore the electron density will remain relatively unaffected. Since the Franck-Condon factors primarily depend on the difference in electron density between different charge states, we have chosen not to take image charges into account in the calculations.

While this effect changes the energies of the orbitals, the shape of the orbitals, and therefore the electron density, on which the Franck-Condon factors primarily depend, will remain relatively unaffected, and we have chosen not to take image charges into account in the calculations.

The omission of the sidearms in the calculation lowers the mass of the molecule, which might explain the discrepancy between the calculation and the measurement for the modes below 10~meV, which involve motions of the whole molecule. Also, the contact geometry in the measurement is unknown, so any mode involving a significant distortion of the gold-sulfur bond is expected to be inaccurate. Like OPV-3, the vibrational spectrum of OPV-5 is less sensitive to the charge state and contact geometry than benzenedithiol and the electron-phonon couplings are smaller (see figure~\ref{fig:eph}c). As in the case of benzenedithiol and OPV-3, the calculation of OPV-5 predicts the intensity of the excitation lines to be much weaker above 30~meV than below. This is also observed in the measurement. The intensities gradually increase for energies up to 30~meV, after which they suddenly drop, a trend also visible in the electron-phonon couplings. For excitations above 80~meV, the low intensities make a quantative comparison between the measurement and the calculation difficult.

Most of the vibrational modes have electron-phonon couplings below 0.1 and are not expected to give rise to extra excitation lines when another vibrational quantum is taken into account. The modes at 17 and 27~meV, with coupling strengths of respectively 0.6 and 0.7, are expected to give rise to excitation lines at 34, 51-54 and possiby 81~meV. These lines are indeed observed in the measurement and the calculation (see figure~\ref{fig:trace}).

It should be emphasized that in figure~\ref{fig:trace}, all visible vibrational excitations, for both the calculation and the measurement, are shown. Comparing the spectrum to Raman and IR spectroscopy data reveals a close match,~\cite{osorio} but the optical spectra predict many more modes not observed in the measurement and calculation. The calculation predicts only a handful visible excitations out of a total of a 129 vibrational modes under 150~meV. Our method is thus able to provide what we might call `selection rules' for vibrational excitations in single-molecule junctions.

\section*{Conclusions}

In conclusion, our results show that the vibrational spectrum of a single-molecule in a junction is sensitive to the contact geometry and charge state, although this influence becomes smaller for larger molecules. Contrary to Raman and IR spectroscopy, calculations can take these influences into account, provide `selection rules' and predict the relative intensity of excitation lines in transport measurements. Our calculations also show that it is necessary to take more than one vibrational quantum into account for small molecules, but that due to decreasing electron-phonon couplings this becomes less important for larger molecules. Finally, our method is computationally efficient. All Franck-Condon and transport calculations have been performed on an HP xw9300 workstation, with the largest calculation (over 54 million Franck-Condon factors in OPV-3 for 250.000 bias and gate points) taking just over four hours.

\section*{Methods}

\textbf{Rate equations.} Our method to calculate stability diagrams is based on the rate equation approach, which is generally used in the sequential tunneling regime.~\cite{beenakker,bonet} The central quantity in this approach is the vector of occupation probabilities $P_{n,\sigma,\nu}$ for states with charge quantum number $n$, spin quantum number $\sigma$ and vibrational quantum number $\nu$. These probabilities change by transitions from one state to the other with rates $R_{n,\sigma,\nu\rightarrow n',\sigma',\nu'}$. The time evolution of the occupation probabilities as a function of these rates is described by the master equation:
\begin{equation} \label{eq:master}
\frac{dP_{n,\sigma,\nu}}{dt}=\sum_{n',\sigma',\nu',\neq n,\sigma,\nu}\left(P_{n',\sigma',\nu'}R_{n',\sigma',\nu'\rightarrow n,\sigma,\nu}-P_{n,\sigma,\nu}R_{n,\sigma,\nu\rightarrow n',\sigma',\nu'}\right)
\end{equation}
The rates are given by Fermi's golden rule:
\begin{equation}
R_{n,\sigma,\nu\rightarrow n',\sigma',\nu'}=\frac{1}{h}\left|\left<n'\sigma'\nu'\left|\hat{H}\right|n\sigma\nu\right>\right|^2\rho_{n',\sigma',\nu'}
\end{equation}
These rates contain contributions from the electronic and nuclear wavefunctions. The electronic contributions are described by the couplings to the leads and the Fermi function, and the nuclear contributions are described by the Franck-Condon factors, which we will discuss below.

At low bias, only two charge states are relevant: the initial state ($n,\sigma',\nu'$), where the level in the bias window (see figure~\ref{fig:device}b) is unoccupied, and the final state ($n+1,\sigma,\nu$), where it is occupied. The rates for these states are:~\cite{koch2,mccarthy}
\begin{widetext}
\begin{align} \label{eq:rates}
R_{n,\sigma',\nu'\rightarrow n+1,\sigma,\nu}^L=&F_{\nu'\nu}\frac{\Gamma_L}{\hbar}f\left(E_{n+1,\sigma,\nu}-E_{n,\sigma',\nu'}-e\beta V_g-e\alpha V_b\right) \\
R_{n,\sigma',\nu'\rightarrow n+1,\sigma,\nu}^R=&F_{\nu'\nu}\frac{\Gamma_R}{\hbar}f\left(E_{n+1,\sigma,\nu}-E_{n,\sigma',\nu'}-e\beta V_g-e\left(\alpha-1\right)V_b\right) \nonumber \\
R_{n+1,\sigma,\nu\rightarrow n,\sigma',\nu'}^L=&F_{\nu'\nu}\frac{\Gamma_L}{\hbar}\left[1-f\left(E_{n+1,\sigma,\nu}-E_{n,\sigma',\nu'}-e\beta V_g-e\alpha V_b\right)\right] \nonumber \\
R_{n+1,\sigma,\nu\rightarrow n,\sigma',\nu'}^R=&F_{\nu'\nu}\frac{\Gamma_R}{\hbar}\left[1-f\left(E_{n+1,\sigma,\nu}-E_{n,\sigma',\nu'}-e\beta V_g-e\left(1-\alpha\right)V_b\right)\right] \nonumber
\end{align}
\end{widetext}
where $F_{\nu'\nu}$ are the Franck-Condon factors and $f$ is the Fermi function. $E_{n+1,\sigma,\nu}-E_{n,\sigma',\nu'}$ is the energy difference between the initial and final state. This difference is composed of the level spacing, the charging energy, the vibrational reorganization energy and the vibrational energy of the states $\nu'$ and $\nu$. For a single degeneracy point, all but the latter of these terms can neglected by choosing a suitable reference point for $V_g$.

With the rate equations in place, we write the master equation~(\ref{eq:master}) in matrix-vector form:
\begin{equation}
\frac{d{\bm P}}{dt}={\bm M}{\bm P}
\end{equation}
where ${\bm P}$ has elements $P_{n,\sigma',\nu'}$ and $P_{n+1,\sigma,\nu}$. ${\bm M}$ is the $2N\times 2N$ rate matrix, where $N$ numbers the vibrational states. Its elements are given by
\begin{equation}
M_{ij}=
\begin{cases}
    -\sum_{\nu=1}^N R_{n,\sigma',i\rightarrow n+1,\sigma,\nu}       & \text{if $i=j$ and $i,j\leq N$,}  \\
    -\sum_{\nu'=1}^N R_{n+1,\sigma,i-N\rightarrow n,\sigma',\nu'}   & \text{if $i=j$ and $i,j>N$,}      \\
    R_{n+1,\sigma,j-N\rightarrow n,\sigma',i}                       & \text{if $i\leq N$ and $j>N$,}    \\
    R_{n,\sigma',j\rightarrow n+1,\sigma,i-N}                       & \text{if $i>N$ and $j\leq N$,}    \\
    0                                                               & \text{otherwise.}
\end{cases}
\end{equation}
To calculate the current we need the stationary occupation probabilities, \emph{i.e.} $\frac{d{\bm P}}{dt}={\bm 0}$, which can be obtained by calculating the null space of ${\bm M}$, with the condition that all elements of ${\bm P}$ are non-negative and $\sum_{n,\sigma,\nu}P_{n,\sigma,\nu}=1$. Once the stationary occupation probabilities and the rates are known, the current can be calculated by summing over the total rate through one of the leads. For the left rate this becomes:~\cite{boese,bonet}
\begin{equation}
I=e\sum_{\nu'=1}^N\sum_{\nu=1}^N\left(P_{n,\sigma',\nu'}R_{n,\sigma',\nu'\rightarrow n+1,\sigma,\nu}^L-P_{n+1,\sigma,\nu}R_{n+1,\sigma,\nu\rightarrow n,\sigma',\nu'}^L\right)
\end{equation}

\textbf{Franck-Condon factors.} The Franck-Condon factors ($F_{\nu'\nu}$) in equation~(\ref{eq:rates}) are a measure of the probability that a tunneling event will be accompanied by a vibrational excitation. When an electron tunnels on or off a molecule, the change in electron density will shift the equilibrium position of the nuclei and possibly cause a transition to a different vibrational excited state. The probability for this to happen is equal to the square of the overlap integral of the vibrational wavefunctions in both charge states.~\cite{wilson} To calculate the overlap integral, the normal coordinates of one charge state have to be expressed in those of the others. This procedure is known as the Duschinsky transformation.~\cite{duschinsky,sando} This transformation yields the Duschinsky rotation matrix and a mass-weighted displacement vector ${\bm k}$. The latter can then be used to calculate the dimensionless electron-phonon couplings $\lambda_i=\sqrt{\frac{\omega_i}{2\hbar}}k_i$.~\cite{mccarthy,ziman}

Two methods are mainly used for calculating the Franck-Condon factors from the Duschinsky transformation and the frequencies. One is the generating function method of Sharp and Rosenstock~\cite{sharp}; the other is the recursion-relation method of Doktorov~\emph{et al.}~\cite{doktorov} We have used the latter as implemented in the two-dimensional array method of Ruhoff and Ratner.~\cite{ruhoff1,ruhoff2} This method returns an $N\times N$ array of Franck-Condon factors, where $N=\binom{\alpha+l}{l}$ is the number of permutations of up to $l$ vibrational quanta over $\alpha$ vibrational modes.

\textbf{Relaxation rates}. Since Franck-Condon factors represent probabilities, the elements of each row and each column of this matrix add up to 1 for $l\rightarrow\infty$. Typically, for a large system with many Franck-Condon factors there are many factors where $F_{\nu'\nu}\ll 1$. This can present numerical difficulties when calculating the stationary occupation probabilities. The rates are proportional to the Franck-Condon factors and if all the rates into and out of a certain level are (very nearly) zero, any occupation of that level will be stationary, resulting in an infinite number of solutions for the stationary occupation probabilities. There are two ways to prevent this from happening. The first is to take intramolecular vibrational relaxation into account. We have implemented a simple relaxation model in which a single relaxation time $\tau$ is used for all states:~\cite{beenakker,koch1}
\begin{align}
\frac{dP_{n,\sigma,\nu}}{dt}=&\sum_{n',\sigma',\nu'\neq n,\sigma,\nu}P_{n',\sigma',\nu'}R_{n',\sigma',\nu'\rightarrow n,\sigma,\nu}-P_{n,\sigma,\nu}R_{n,\sigma,\nu\rightarrow n',\sigma',\nu'} \nonumber \\
&-\frac{1}{\tau}\left(P_{n,\sigma,\nu}-P_{n,\sigma,\nu}^{eq}\sum_{\nu''}P_{n,\sigma,\nu''}\right)
\end{align}
where
\begin{equation}
P_{n,\sigma,\nu}^{eq}=\frac{e^{-\frac{E_{n,\sigma,\nu}}{kT}}}{\sum_{\nu'}e^{-\frac{E_{n,\sigma,\nu'}}{kT}}}
\end{equation}
is the equilibrium population according to the Boltzmann distribution. This term can be included in the rate matrix by adding the relaxation matrix:
\begin{equation}
M_{ij}^R=\frac{1}{\tau}\times
\begin{cases}
    P_{n,\sigma',i}^{eq}-1      & \text{if $i=j$ and $i,j\leq N$,}      \\
    P_{n,\sigma',i}^{eq}        & \text{if $i\neq j$ and $i,j\leq N$,}  \\
    P_{n+1,\sigma,i-N}^{eq}-1   & \text{if $i=j$ and $i,j>N$,}          \\
    P_{n+1,\sigma,i-N}^{eq}     & \text{if $i\neq j$ and $i,j>N$,}      \\
    0                           & \text{otherwise.}
\end{cases}
\end{equation}
For sufficiently small relaxation times, the previously mentioned stationary states will decay to the ground state and there will be only one solution.

\textbf{Iterative solution.} The second approach is to calculate the null space iteratively by starting from the equilibrium population. Since both the equilibrium population of higher energy states and the rates to those states are zero, they will never become populated and the method will converge to the physically correct solution. This approach has several additional benefits. The rate matrix for most realistic systems will be sparse and an iterative method can make use of this and scale better than a direct method. Also, the stationary population at neighbouring bias and gate points will be the same unless a new state becomes available, so by using the previous population as a starting point, most points will only need a single iteration to converge. We have implemented a direct method using singular value decomposition and an iterative method using a Jacobi preconditioned biconjugate gradient method. The second implementation is generally several orders of magnitude faster. It turns out that implementing a combination of both approaches yields a one-dimensional null space of the rate matrix, even for larger molecules with several vibrational quanta.

\subsection*{Acknowledgment}
We thank M. Galperin and S. Yeganeh for discussions. Financial support was obtained from Stichting FOM (project 86), from the EU FP7 programme under the grant agreement ``SINGLE'', and from the Division of Chemistry and the Office of International Science and Engineering of the NSF in the US. This work was also sponsored by the Stichting Nationale Computerfaciliteiten (National Computing Facilities Foundation, NCF, project mp-06-111) for the use of supercomputer facilities, with financial support from the Nederlandse Organisatie voor Wetenschappelijk Onderzoek (Netherlands Organization for Scientific Research, NWO).


\begin{thebibliography}{99}

\bibitem{stipe} Stipe, B. C.; Rezaei, M. A.; Ho, W. Single-Molecule Vibrational Spectroscopy and Microscopy \textit{Science} \textbf{1998}, \textit{280}, 1732--1735.
\bibitem{smit} Smit, R. H. M.; Noat, Y.; Untiedt, C.; Lang, N. D.; van Hemert, M. C.; van Ruitenbeek, J. M. Measurement of the Conductance of a Hydrogen Molecule \textit{Nature} \textbf{2002}, \textit{419}, 906--909.
\bibitem{parks} Parks, J. J.; Champagne, A. R.; Hutchison, G. R.; Fores-Torres, S.; Abru\~na, H. D.; Ralph, D. C. Tuning the Kondo Effect with a Mechanically Controllable Break Junction \textit{Phys. Rev. Lett.} \textbf{2007}, \textit{99}, 026601.
\bibitem{park} Park, H.; Park, J.; Lim, A. K. L.; Anderson, E. H.; Alivisatos, A. P.; McEuen, P. L. Nanomechanical Oscillations in a Single-C$_{60}$ Transistor \textit{Nature} \textbf{2000}, \textit{407}, 57--60.
\bibitem{yu} Yu, L. H.; Keane, Z. K.; Ciszek, J. W.; Cheng, L.; Stewart, M. P.; Tour, J. M.; Natelson, D. Inelastic Electron Tunneling via Molecular Vibrations in Single-Molecule Transistors \textit{Phys. Rev. Lett.} \textbf{2004}, \textit{93}, 266802.
\bibitem{pasupathy} Pasuapthy, A. N.; Park, J.; Chang, C.; Soldatov, A. V.; Lebedkin, S.; Bialczak, R. C.; Grose, J. E.; Donev, L. A. K.; Sethna, J. P.; Ralph, D. C. \emph{et al.} Vibration-Assisted Electron Tunneling in C$_{140}$ Transistors \textit{Nano Lett.} \textbf{2005}, \textit{5}, 203--207.
\bibitem{osorio} Osorio, E. A.; O'Neill, K.; Stuhr-Hansen, N.; Nielsen, O. F.; Bj\o rnholm, T.; van der Zant, H. S. J. Addition Energies and Vibrational Fine Structure Measured in Electromigrated Single-Molecule Junctions Based on an Oligophenylenevinylene Derivative \textit{Adv. Mater.} \textbf{2007}, \textit{19}, 281--285.
\bibitem{boese} Boese, D.; Schoeller, H.; Influence of Nanomechanical Properties on Single-Electron Tunneling: A Vibrating Single-Electron Transistor \textit{Europhys. Lett.} \textbf{2001}, \textit{54}, 668--674.
\bibitem{braig} Braig, S.; Flensberg, K. Vibrational Sidebands and Dissipative Tunneling in Molecular Transistors \textit{Phys. Rev. B} \textbf{2003}, \textit{68}, 205324.
\bibitem{mitra} Mitra, A.; Aleiner, I.; Millis, A. J. Phonon Effects in Molecular Transistors: Quantal and Classical Treatment \textit{Phys. Rev. B} \textbf{2004}, \textit{69}, 245302.
\bibitem{wegewijs} Wegewijs, M. R.; Nowack, K. C. Nuclear Wavefunction Interference in Single-Molecule Electron Transport \textit{New J. of Phys.} \textbf{2005}, \textit{7}, 239.
\bibitem{koch1} Koch, J.; von Oppen, F.; Oreg, Y.; Sela, E. Thermopower of Single-Molecule Devices \textit{Phys. Rev. B} \textbf{2004}, \textit{70}, 195107.
\bibitem{koch2} Koch, J.; von Oppen, F.; Andreev, A. V. Theory of the Franck-Condon Blockade Regime \textit{Phys. Rev. B} \textbf{2006}, \textit{74}, 205438.
\bibitem{chang} Chang, C. T.; Sethna, J. P.; Pasupathy, A. N.; Park, J.; Ralph, D. C.; McEuen, P. L. Phonons and Conduction in Molecular Quantum Dots: Density Functional Calculations of Franck-Condon Emission Rates for Bifullerenes in External Fields \textit{Phys. Rev. B} \textbf{2007}, \textit{76}, 045435.
\bibitem{thijssen} Thijssen, J. M.; van der Zant, H. S. J. Charge Transport and Single-Electron Effects in Nanoscale Systems \textit{Phys. Stat. Sol. b} \textbf{2008}, doi:10.1002/pssb.200743470.
\bibitem{adf1} Fonseca Guerra, C.; Snijders, J. G.; te Velde, G.; Baerends, E. J. Towards an Order-N DFT Method \textit{Theor. Chem. Acc.} \textbf{1998}, \textit{99}, 391--403.
\bibitem{adf2} te Velde, G.; Bickelhaupt, F. M.; van Gisbergen, S. J. A.; Fonseca Guerra, C.; Baerends, E. J.; Snijders, J. G.; Ziegler, T. Chemistry with ADF \textit{J. Comput. Chem.} \textbf{2001}, \textit{22}, 931--967.
\bibitem{adf3} Wolff, S. K. Analytical Second Derivatives in the Amsterdam Density Functional Package \textit{Int. J. Quantum Chem.} \textbf{2005}, \textit{104}, 645--659.
\bibitem{kubatkin} Kubatkin, S.; Danilov, A.; Hjort, M.; Cornil, J.; Br\'{e}das, J. L.; Stuhr-Hansen, N.; Hedeg\aa rd, P.; Bj\o rnholm, T. Single-Electron Transistor of a Single Organic Molecule with Access to Several Redox States \textit{Nature} \textbf{2003}, \textit{425}, 698--701.
\bibitem{beenakker}  Beenakker, C. J. B. Theory of Coulomb-Blockade Oscillations in the Conductance of a Quantum Dot \textit{Phys. Rev. B} \textbf{1991}, \textit{44}, 1646--1656.
\bibitem{bonet} Bonet, E.; Desmukh, M. M.; Ralph, D. C. Solving Rate Equations for Electron Tunneling via Discrete Quantum States \textit{Phys. Rev. B} \textbf{2002}, \textit{65}, 045317.
\bibitem{mccarthy} McCarthy, K. D.; Prokov'ev, N.; Tuominen, M. T. Incoherent Dynamics of Vibrating Single-Molecule Transistors \textit{Phys. Rev. B} \textbf{2003}, \textit{67}, 245415.
\bibitem{wilson} Wilson, E. B.; Decius, J. C.; Cross, P. C. \textit{Molecular Vibrations}; McGraw-Hill: New York, 1955.
\bibitem{duschinsky} Duschinsky, F. Meaning of the Electronic Spectrum of Polyatomic Molecules. I. The Franck-Condon Principle \textit{Acta Physicochim. URSS} \textbf{1937}, \textit{7}, 551--566.
\bibitem{sando} Sando, G. M.; Spears, K. G. Ab Initio Computation of the Duschinsky Mixing of Vibrations and Nonlinear Effects \textit{J. Phys. Chem. A} \textbf{2001}, \textit{105}, 5326--5333.
\bibitem{ziman} Ziman, J. M. \textit{Electrons and Phonons}; Oxford University Press: New York, 2001.
\bibitem{sharp} Sharp, T. E.; Rosenstock, H. M. Franck-Condon Factors for Polyatomic Molecules \textit{J. Chem. Phys.} \textbf{1964}, \textit{41}, 3453--3463.
\bibitem{doktorov} Doktorov, E. V.; Malkin, I. A.; Man'ko, V. I. Dynamical Symmetry of Vibronic Transitions in Polyatomic Molecules and the Franck-Condon Principle \textit{J. Mol. Spectrosc.} \textbf{1975}, \textit{56}, 1--20.
\bibitem{ruhoff1} Ruhoff, P. T. Recursion Relations for Multi-Dimensional Franck-Condon Overlap Integrals \textit{Chem. Phys.} \textbf{1994}, \textit{186}, 335--374.
\bibitem{ruhoff2} Ruhoff, P. T.; Ratner, M. A. Algorithms for Computing Franck-Condon Overlap Integrals \textit{Int. J. Quantum Chem.} \textbf{2000}, \textit{77}, 383--392.

\end{thebibliography}
\end{document}